\documentclass[aps,prd,preprintnumbers,showpacs]{revtex4}
\usepackage[dvips]{graphicx}
\begin{document}

%
%

\preprint{Nisho-2-2009}
\title{Pair Creation of Massless Fermions in Electric Flux Tube}
\author{Aiichi Iwazaki}
\address{International Economics and Politics, Nishogakusha University,\\ 
Ohi Kashiwa Chiba  277-8585, Japan.}   
\date{May 13, 2009}
\begin{abstract}
Using chiral anomaly,
we discuss the pair creation of massless fermions in an electric flux tube $\vec{E}$
under homogeneous magnetic field $\vec{B}$ parallel to $\vec{E}$.
The tube is axial symmetric and 
infinitely long in longitudinal direction.
In the limit
$B\gg E$,
we can analytically obtain the spatial and temporal behaviors of the electric field and 
azimuthal magnetic field 
generated by the produced fermions.
We find that the life time $t_c$ of the electric field is shorter as the width of the tube is narrower. 
Applying it to the glasma in high-energy heavy-ion collisions,
we find that color electric field decays fast such
as $t_c\simeq Q_s^{-1}$ with saturation momentum $Q_s$.
\end{abstract}
\hspace*{0.3cm}
\pacs{12.38.-t, 24.85.+p, 12.38.Mh, 25.75.-q  \\
Schwinger mechanism, Chiral Anomaly, Color Glass Condensate}
\hspace*{1cm}

\maketitle

Pair creation of charged particles in a classical strong electric field 
has been theoretically discussed\cite{tanji} for long time since the
discovery of Klain paradox,
although such phenomena have not yet been observed. 
The particle creation is known as Schwinger mechanism\cite{schwinger}.
Here, we are addressed with the problem in a quite different way from
previous ones and apply our results to 
color gauge fields (glasma) produced in high-energy heavy-ion collisions.

The glasma
have recently received much attention. 
The gauge fields are classical longitudinal color electric and magnetic fields.
It is expected that
the decay of the glasma leads to thermalized quark gluon plasma (QGP)
observed in recent RHIC experiments. 
It is a possible mechanism that the color electric field decays with the pair creation of quarks.
Indeed, the color electric field has been phenomenologically assumed in previous papers\cite{iwazaki} 
to be generated in the collisions and to lead to QGP owing to the pair creation of quarks.  
On the other hand, the generation of the classical gauge fields 
has recently been argued on the basis of a fairly reliable
effective theory of QCD at high energy, that is, a model of color glass condensate (CGC)\cite{colorglass}.

According to a model of the CGC, 
the glasma is homogeneous in the longitudinal
direction and inhomogeneous in the transverse directions. Hence,
they form color electric and magnetic flux tubes extending in the longitudinal direction. 
In our previous papers\cite{iwa,itakura,hii}, 
we have discussed a possible mechanism for the decay of the color magnetic field
and have reproduced the instability of the glasma
observed in numerical simulations\cite{venugopalan,berges}. 
In the discussion, only classical gluon's dynamics has been considered.
On the othe hand, taking account of quark's dynamics such as quark pair creation in color electric field,
we have discussed\cite{iwa2}
the decay of homogeneous color electric field.
In particular,
we have analytically obtained
the temporal evolution of the electric field as well as the number density of the quarks.   
In the analysis,
we have not explicitly used the wavefunctions of the quarks in the electric field.
Instead, we have simply used the formula of chiral anomaly, which already involves the quantum effects of pair creation
and annihilation.

In the present paper we generalize our previous discussion\cite{iwa2} to electric field which forms an axial symmetric flux tube
infinitely extending in the longitudinal direction denoted by the coordinate $z$.
We discuss the decay of the field which loses its energy owing to 
the acceleration of produced fermions.
These phenomena occur in cylindrically symmetric way
under the effect of longitudinal strong magnetic field $B\gg E$. 
Therefore,
we need to discuss an azimuthal magnetic field generated by the current of the fermions.
Indeed, we can analytically obtain the spatial and temporal behaviors of the azimuthal magnetic field and
the electric field after the axial symmetric electric flux tube is switched on.
The fields depend on a cylindrical coordinate $r=\sqrt{x^2+y^2}$ as well as time 
coordinate $t$. As far as we know, it is first to obtain
the analytic formulation of the azimuthal magnetic field 
arising in the axial symmetric electric field.


Here we mainly discuss the pair creation of electrons and positrons for simplicity.
The generalization to QCD is straightforward.
Before discussing the pair creation in the electric flux tube,
we first give a brief review of the pair creation in homogeneous electric and magnetic fields.
We assume that the fermions are massless and  
that the fields $\vec{B}$ and $\vec{E}$ are spatially homogeneous and parallel (or antiparallel) with each other.
They are oriented in the $z$ direction.
Then, the energies of electrons with charge $-e<0$ and positrons with charge $e>0$ 
under the magnetic field $B=|\vec{B}|$ are given by

\begin{equation}
\label{1}
E_N=\sqrt{p_z^2+2NeB} \quad \mbox{(parallel)} \quad \mbox{and} \quad E_N=\sqrt{p_z^2+eB(2N+1)} \quad \mbox{(antiparallel)},
\end{equation}
where $p_z$ denotes momentum in the $z$ direction and 
integer $N\ge 0$ does Landau level. The term of
``parallel" (``antiparallel") implies magnetic moment parallel (antiparallel) to $\vec{B}$. 
The magnetic moment of electrons (positrons) is antiparallel (parallel) to their spin.
Thus,
electrons (positrons) with spin antiparallel (parallel) to $\vec{B}$ can have zero energy states
in the lowest Landau level; their energy spectrum is given by $E_{N=0}=|p_z|\ge 0$. 
On the other hand, the other states cannot be zero energy states; their energy spectra are given
by $\sqrt{p_z^2+eBM}\ge eBM\ge eB$ with positive integer $M$.
In a sense they are ``massive" states whose masses increase with $B$.  

When the electric field is switched on,
the pair creation of electrons and positrons with the energies $E_N$
occurs.
However, it is not probable that any states are produced 
with an equal production rate. Indeed, high energy states are
hard to be produced while low energy states are easy to be produced.
Thus,
the ``massive" states are hard to be produced by weak electric field $E \ll B$.
In particular, they cannot be produced in the limit of $B \to \infty $.
Only products in the limit are the pairs of electrons and positrons in zero energy states ( $E_{N=0}=|p_z|=0$ ).

Note that owing to the Fermi statistics
only fermions with $p_z=0$ are produced,
since the other states with $p_z\neq 0$ have already been occupied.
After the production of the fermions with $p_z=0$, their momenta
increase with time, $p_z(t)=\pm e\int^t dt' E$, owing to the acceleration by the electric field.
Hence we obtain the momentum distribution $\tilde{n}(p_z)\propto \theta(p_F(t)-p_z)\theta(p_z)$ 
for positrons with charge $e>0$ and $\tilde{n}(p_z)\propto \theta(p_F(t)+p_z)\theta(-p_z)$
for electrons with charge $-e<0$. 
The Fermi momentum $p_F$ is given by $p_F(t)=e\int_0^tdt'E(t')$.
Here we have assumed that the electric field is parallel to $\vec{B}$
and is switched on at $t=0$.  
(When $\vec{E}$ is antiparallel to $\vec{B}$, the momentum of the particles 
increases in the different direction from the case of $\vec{E}$ being parallel $\vec{B}$.)

We note that electrons move to the direction antiparallel to $\vec{E}$ while
positrons move to the direction parallel to $\vec{E}$.
Therefore, both electrons and positrons created by the electric field have right handed helicity
when $\vec{E}$ being parallel $\vec{B}$, while they have left handed helicity when $\vec{E}$ being antiparallel $\vec{B}$. 


In our discussion of the pair creation,
the key point is to use the equation of chiral anomaly,

\begin{equation}
\label{chiral}
 \partial_t (n_R-n_L)=\frac{e^2}{4\pi^2}E(t)B
\end{equation}
where $n_R$ ( $n_L$ ) denotes number density of right ( left ) handed chiral fermions; 
$n_{R,L}=\langle\bar{\Psi}\gamma_0(1\pm \gamma_5)\Psi\rangle/2$
in which the expectation value is taken by using a state of electrons and positrons created
in the electric field. Here, we have used spatial homogeneity of the chiral current $\vec{j}_5$, that is, $\rm{div}\vec{j}_5=0$.

In accordance with the chiral anomaly, 
the rate of chirality change is given by the product of $\vec{E}$ and $\vec{B}$.
It is important to note that only particles with right ( left ) handed helicity
are produced in the limit $B\gg E$ when $\vec{E}$ is parallel to $\vec{B}$ ($\vec{E}$ being antiparallel to $\vec{B}$.) 
Since the number density $n$ of electrons
is the same as that of positrons, $n_R=2n$ and $n_L=0$ when $\vec{E}$ being parallel to $\vec{B}$, while $n_R=0$ and $n_L=2n$ 
when $\vec{E}$ being antiparallel to $\vec{B}$.
Thus, we find that the temporal evolution of the number density
is determined by the anomaly equation
$2\partial_t n(t)=e^2|E(t)|B/4\pi^2$.


After the production of the particles, the electric field gradually loses its energy 
because it accelerates the particles.
The energy of the electric field is transformed into the
energies $\epsilon$ of the particles, so that the energies of the particles increase,

\begin{equation}
\partial_t\Bigl(\epsilon(t) +\frac{1}{2}E(t)^2\Bigr)=0
\end{equation} 
where we have neglected the contribution of magnetic field induced by the electric current of electrons and positrons.
Later we take into account the contribution when we discuss pair production in an electric flux tube.

The energy density $\epsilon(t)$ of the particles are given
such as $\epsilon(t)=2n(t)p_F(t)/2=n(t)p_F(t)$.
This originates from the fact that the momentum distribution $\tilde{n}(p_z)$ ( $n(t)\equiv \pm\int_0^{\pm\infty} dp_z \tilde{n}(p_z)$ )
is given by $\tilde{n}(p_z)=n_0\theta (p_F(t)-|p_z|)\theta(|p_z|)$ as we have discussed.
For example, the energy density of electrons is given such that $\int_{-\infty}^0 dp_z \tilde{n}(p_z)|p_z|=np_F/2$.

Now we have three equations to solve for obtaining the pair production rate $\partial_t n(t)$, etc.,

\begin{equation}
\label{tot}
2\partial_t n(t)=\frac{e^2}{4\pi^2}E(t)B, \quad \partial_t(\epsilon(t) +\frac{1}{2}E(t)^2)=0, \quad \mbox{and}
\quad \epsilon(t)=n(t)p_F(t)
\end{equation}
with $p_F(t)=\int_0^t dt' eE(t')$.
It is easy to solve the eq(\ref{tot}) with initial conditions $E(t=0)=E_0>0$ and $n(t=0)=0$ by
assuming magnetic field $B$ independent on $t$,

\begin{equation}
\label{result}
E(t)=E_0\cos(\sqrt{\frac{\alpha eB}{\pi}}\,t) \quad \mbox{and} \quad 
n(t)=\frac{\alpha E_0B|\sin(\sqrt{\frac{\alpha eB}{\pi}}\,t)|}{\pi \sqrt{\frac{\alpha eB}{\pi}}}
\end{equation}
with $\alpha=e^2/4\pi$, where we have taken account of the number density being positive semi-definite.
We find that the life time of the electric field, which may be defined as $E(t_c)=0$,
is given by $t_c=\frac{\pi}{2}(\sqrt{\frac{\alpha eB}{\pi}})^{-1}$.

The formula holds rigorously in the limit $B\gg E_0$ since the particles with masses on the order of $B$ are
suppressed in the production.
In Fig.\,1 we have shown the behaviors $E(t)=E_0\cos(\pi \,t/2t_c)$ and $n(t)=\frac{2E_0\alpha Bt_c}{\pi^2}|\sin(\pi\,t/2t_c)|$ 
with $E_0=1$, $t_c=1$ and $\alpha\simeq 1/137$.
When the electric field is switched on at $t=0$, the pair creation begins to occur, so that the number density 
of electrons and positrons increases.
Owing to the acceleration of the particles by the electric field, 
the energy of the electric field gradually decreases and eventually vanishes at $t=t_c$.
At the same time the number density takes the maximum value. Then, the direction of the electric field is flipped
and it becomes strong with time. On the other hand, the number density decreases after $t=t_c$. 
This decrease is caused by the pair annihilation of electrons and positrons. Since the direction of the electric field is flipped
after $t_c$, the direction of the particle acceleration is also flipped.
Consequently, electrons and positrons are moved to overlap with each other so that the pair annihilation
may occur to make the number density decrease. Eventually, $n$ vanishes at $t=2t_c$. 
Then, the pair creation of left-handed particles begins to occur 
since $\vec{E}$ is anti-parallel to $\vec{B}$ after $t=2t_c$.
Therefore, the oscillation of $E(t)$ and $n(t)$ arises. 

\begin{figure}[t]
\begin{minipage}{.47\textwidth}
\includegraphics[width=6.9cm,clip]{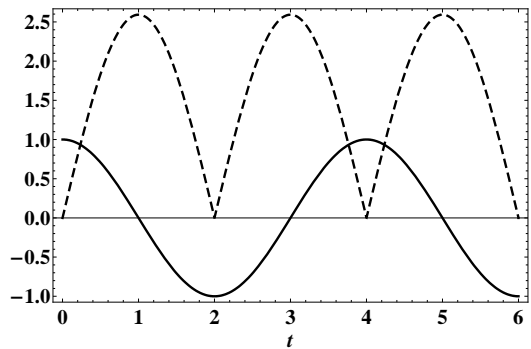}
\label{fig:growth-rate3}
\caption{electric field $E(t)$ (solid line) and number density $n(t)$ (dashed line)}
\end{minipage}
\end{figure} 

We wish to make a comment that although the formula of the chiral anomaly involves all of the quantum effects, 
the mechanism why the number density increases
is not explicit in the formula. However,
we understand that the increase or decrease of the number density originates with the pair creation or annihilation.
This is because the results shown in this paper have agreed with those obtained by
calculations\cite{tanji} explicitly including the quantum effects of the pair creation or annihilation.


\vspace*{0.2cm}
Now we discuss the pair production in an axial symmetric flux tube of electric field, namely the field
extending infinitely in the longitudinal direction, but has finite width $R$ in the transverse directions.
In this case the quantities of $E$, $n$ and $p_F$ depend on the
cylindrical coordinate $r$ and time coordinate $t$. 
( Since the wavefunctions in the Lowest Landau level are given by $z^m\exp(-eB|z|^2/4)$ with $z=x+iy$ and integer $m\ge 0$,
the charged particles can be located within the flux tube for sufficiently large $B$ such as $R\gg 1/\sqrt{eB}$. )
Thus, the formula of the chiral anomaly takes
the same form as the one in eq(\ref{tot}), although the electric field $E(r,t)$ and the number density $n(r,t)$
depend on $r$ and $t$. ( Owing to the axial symmetry and homogeneity in the longitudinal direction,
we may take $div\vec{j}_5=0$. )
Similarly, the energy density $\epsilon$ takes the form $\epsilon(r,t)=n(r,t)p_F(r,t)$ as eq(\ref{tot}),
in which the Fermi momentum is given by $p_F(r,t)=\int_0^{t}dt'eE(r,t')$.
We also take into account Maxwell equations such as $\partial_t B_{\theta}(r,t)=\partial_r E(r,t)$ and
$\partial_t E(r,t)=\frac{\partial_r(rB_{\theta}(r,t))}{r}-J(r,t)$ where $J(r,t)$ represents the current of electrons and positrons
flowing in the $z$ direction. $B_{\theta}(r,t)$ denotes azimuthal magnetic field generated by the current.
Therefore, we have the following equations to solve,

\begin{equation}
2\partial_t n(r,t)=\frac{e^2}{4\pi^2}E(r,t)B, \quad \epsilon(t)=n(t)p_F(t), 
\end{equation}
and Maxwell equations,
\begin{equation}
\label{max}
\partial_t B_{\theta}(r,t)=\partial_r E(r,t),
\quad \partial_t E(r,t)=\frac{\partial_r(rB_{\theta}(r,t))}{r}-J(r,t).
\end{equation}

In order to solve the equations we must represent the current $J(r,t)$ in terms of $n(r,t)$ and $p_F(r,t)$.
It can be done by imposing the condition of energy conservation,

\begin{equation}
\partial_t\int d^3x \Bigl(\frac{E^2(r,t)+B_{\theta}^2(r,t)}{2}+\epsilon(r,t)\Bigr)=\int d^3x \Bigl(E(r,t)\partial_t E(r,t)
+B_{\theta}(r,t)\partial_tB_{\theta}(r,t)+\partial_t\epsilon(r,t)\Bigr)=0.
\end{equation}
The condition implies that the energy of the electric field is transmitted into the energies of
the azimuthal magnetic field $B_{\theta}$ and the produced electrons and positrons.
The particles are accelerated by the electric field so that 
the field induces the electric current $J$, which produces
the field $B_{\theta}$.

Using the Maxwell equations in eq(\ref{max}), we rewrite the condition such that

\begin{equation}
\int d^3x \Bigl(E(\partial_tE-\frac{\partial_r(rB_{\theta})}{r})+\partial_t\epsilon \Bigr)=
\int d^3x(-JE+\partial_t\epsilon)=\int d^3x E(en+\frac{e^2}{8\pi^2}B p_F-J)=0,
\end{equation}
where we have performed the partial integration in $r$.
The energy conservation must hold for any initial conditions of $E(r,t=0)=E_0(r)$.
Therefore, we obtain 

\begin{equation}
\label{J}
J(r,t)=en(r,t)+\frac{e^2}{8\pi^2}p_F(r,t)B=2en(r,t),
\end{equation}
since $\frac{e^2}{8\pi^2}p_F(r,t)B=\frac{e^2}{8\pi^2}\int_0^t dt' eE(r,t')B=en(r,t)$ with $n(r,t=0)=0$.
It states that both of electrons with charge density $-en$ and positrons with $en$ 
constitute the electric current $J$ since the velocity of electrons is $-1$, while that of positrons is $+1$
in the unit of light velocity $c=1$. Obviously,
the expression $J=2en$ is very natural.

Using $J(r,t)$ in eq(\ref{J}) as well as equations in eq(\ref{tot}), we derive the equation of motion of the electric field,

\begin{equation}
\label{E}
\partial_t^2E(r,t)=(\partial_r^2+\frac{1}{r}\partial_r-\frac{e^3B}{4\pi^2})E(r,t),
\end{equation}
where we have used the formula $\partial_tJ(r,t)=\frac{e^3}{4\pi^2}BE(r,t)$.

We solve the equation(\ref{E}) with the initial conditions $n(r,t=0)=0$, $B_{\theta}(r,t=0)=0$ and $E(r,t=0)=E_0\exp(-r^2/R^2)$.
The initial conditions implies that the electric field forming flux tube with radius $R$ is switched on at $t=0$,
when any particles and magnetic field $B_{\theta}$ are absent. 
Noting that $\partial_t E(r,t=0)=0$ derived from Maxwell equations (\ref{max}) as well as the initial conditions,
we first obtain the solution $E$, and then we derive $n$ and $B_{\theta}$, 

\begin{eqnarray}
\label{sol1}
E(r,t)&=&\frac{E_0R^2}{2}\int_0^{\infty}kdk \cos(t\sqrt{k^2+m^2})J_0(kr)\exp(-k^2R^2/4), \\
\label{sol2}
n(r,t)&=&\frac{e^2E_0R^2}{16\pi^2}\Bigl|\int_0^{\infty} kdk \frac{\sin(t\sqrt{k^2+m^2})}{\sqrt{k^2+m^2}}J_0(kr)\exp(-k^2R^2/4)\Bigr|, \\
\label{sol3}
B_{\theta}(r,t)&=&-\frac{E_0R^2}{2}\int_0^{\infty} kdk \frac{\sin(t\sqrt{k^2+m^2})}{\sqrt{k^2+m^2}}J_1(kr)\exp(-k^2R^2/4),
\end{eqnarray}
with $m^2\equiv\frac{e^3}{4\pi^2}B$,
where $J_i(kr)$ denotes Bessel functions. We have used the fact that the number density $n$ must be positive. 
Taking the limit $R\to \infty$ in eq(\ref{sol1})$\sim$ eq(\ref{sol3}),
we find that the solutions in eq(\ref{sol1}) $\sim$ eq(\ref{sol3}) are reduced to the previous solutions in eq(\ref{result}),
which are obtained under the condition that 
the electric field is homogeneous, in other words, the flux tube with infinite width.

We have shown the temporal and spatial behaviors of electric field $E(r,t)$ in Fig.2, number density $n(r,t)$ in Fig.3,
and azimuthal magnetic field $B_{\theta}$ in Fig.4, respectively, where the parameters $m=1$, $E_0R^2/2=1$, $e^2/8\pi^2=1$ and $R=1$ have been used;
the scale $r=10$ in the figures corresponds to $R$. Small and large dots represent the behaviors at $t=0.2$ and $t=0.6$, respectively. 
We find that the electric field $E(r,t)$ vanishes faster as $r$ is smaller since $E(r,t=0)$ is stronger as $r$ is smaller.
The strong electric field produces the particles so much that the number density of the particles is larger as $r$ is smaller. 
We also find from the solutions (\ref{sol1}) that the life time of the electric field with finite $R$ is
shorter than that of the field with $R=\infty$. ( The life time may be defined roughly as $E(r=R,t=t_c)=0$,
although the field oscillates with time. )
This is because typical momentum dominating in the $k$ integration is on the order of $R^{-1}$. 
Thus, the life time is approximately given by $t_c\simeq (\sqrt{m^2+R^{-2}})^{-1}$.

\begin{figure}[t]
\begin{minipage}{.47\textwidth}
\includegraphics[width=7.1cm,clip]{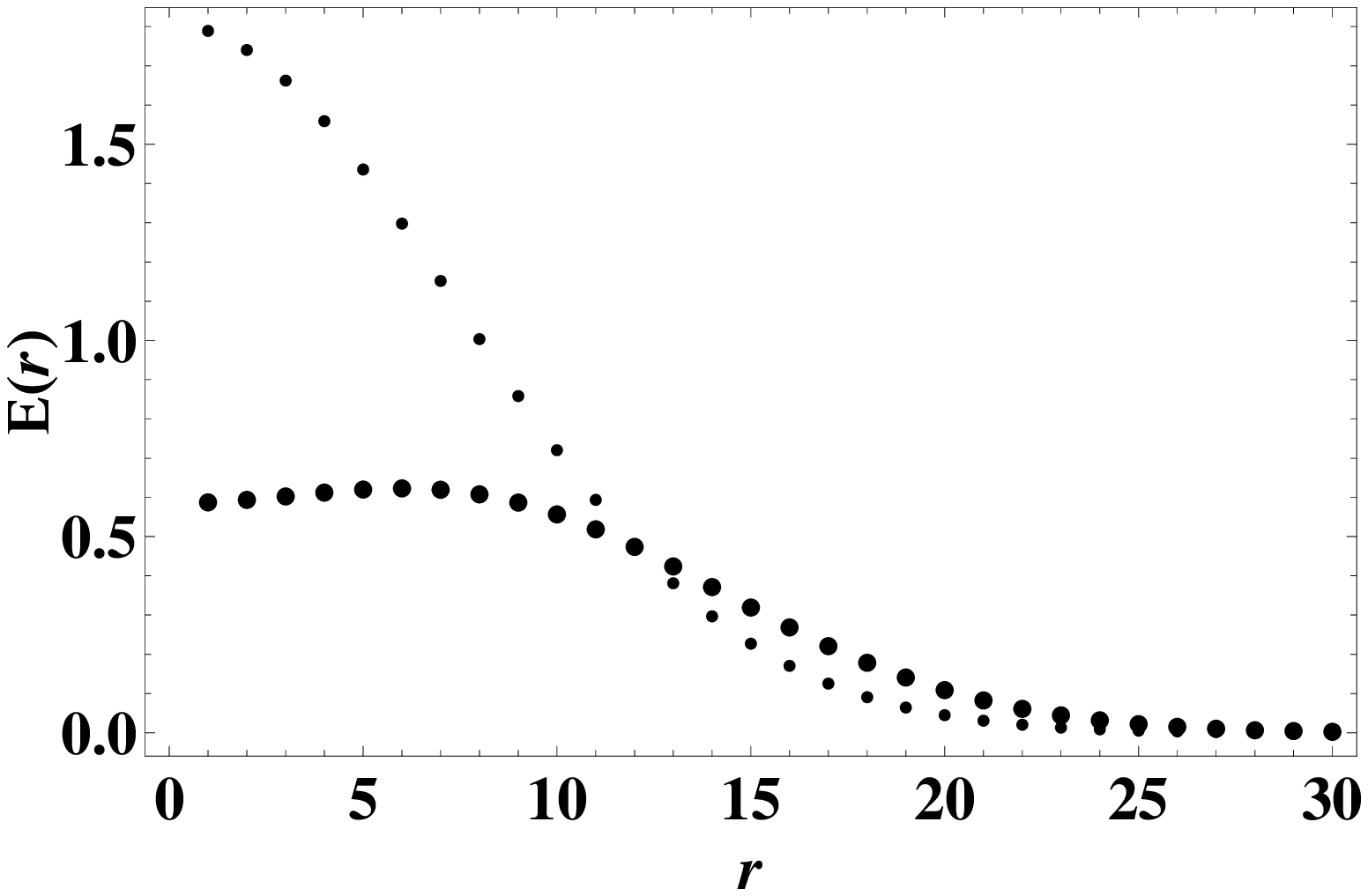}
\label{fig:E}
\caption{electric field $E(r)$ at $t=0.2$ (small dots) \\ and $E(r)$ at $t=0.6$ (large dots)}
\end{minipage}
\hfill
\begin{minipage}{.47\textwidth}
\includegraphics[width=7.1cm,clip]{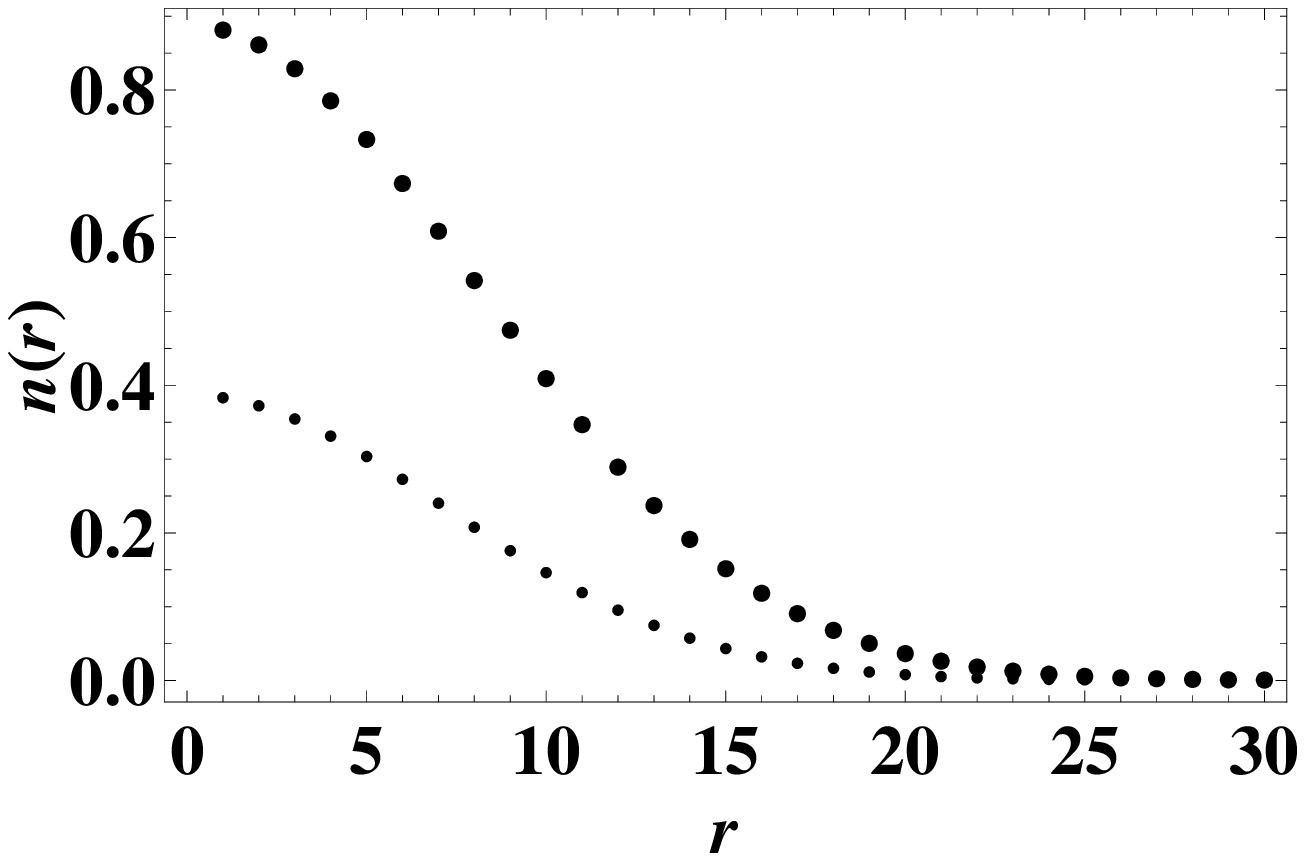}
\label{fig:n}
\caption{number density $n(r)$ at $t=0.2$ (small dots) \\ and $n(r)$ at $t=0.6$ (large dots)}
\end{minipage}
\end{figure}

\begin{figure}[t]
\begin{minipage}{.47\textwidth}
\includegraphics[width=7.1cm,clip]{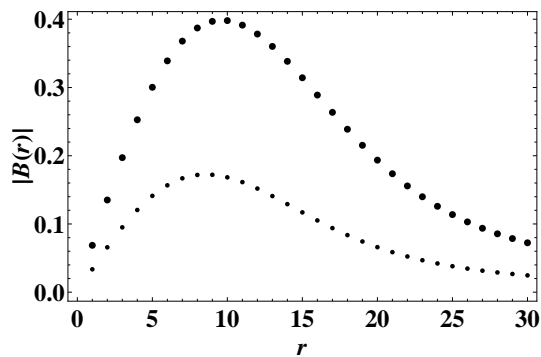}
\label{fig:B}
\caption{azimuthal magnetic field $B_{\theta}(r)$ at $t=0.2$ \\(small dots) and $B_{\theta}(r)$ at $t=0.6$ (large dots)}
\end{minipage}
\end{figure}

As we can see, the quantities such as $E(r,t)$, $n(r,t)$, and $B_{\theta}(r,t)$ oscillate with time.
In particular, electric field $E(r,t)$ never disappear by completely losing its energy.
This originates with the fact that the produced particles are almost free
and never interact with external so that they never lose their energies and momenta.
Indeed, we have used the momentum distribution of the free fermions.
On the other hand, when we take into account their interactions with external heat bath and 
use fermionic momentum distributions at finite temperature,
the electric field would decay and disappear
because the current $J$ would vanish owing to the dissipation of the particle's momenta
in the heat bath. 

Distinguished point in our analysis is that simply when we adopt an appropriate momentum distribution
including the effects of particle interactions, we can discuss 
the disappearance of electric field using the chiral anomaly in the limit $B\gg E$.
We do not need to explicitly obtain the wavefuncions of the interacting particles.
For this simplicity, we are able to explore a mechanism for the disappearance of the electric field.

It is interesting to see from eq(\ref{E}) that the electric field $\vec{E}$ parallel to the background magnetic field $\vec{B}$ possesses
an effective mass $m=\sqrt{\alpha eB/\pi}$ with $\alpha=e^2/4\pi$ owing to the pair creation of electrons and positrons. 
This result holds only for $B\gg E$ and the pair production of massless fermions. 
When the fermions have a mass ( indeed, electrons have small mass ), we need to modify the equation of the chiral anomaly 
and momentum distributions of the fermions. However, the modification of our results
would be minor when the mass is much less than $B$.
We will discuss the effect of fermion mass on the pair creation in future publications.

Finally we wish to make a comment on the decay of the glasma produced in high-energy heavy-ion collisions.
The color electric and magnetic fields generated in the early stage of the collisions form flux tubes.
Although we have assumed that magnetic field is not a tube-like but homogeneous,
we may apply our results to the analysis of the glasma decay, in particular, the decay of the color electric field. 
Then, as we found, the life time $t_c$ of electric field is approximately given by $(\sqrt{R^{-2}+m^2})^{-1}$.
In the glasma, $m^2$ is of order $\alpha_s gB$ where $gB \simeq Q_s^2$ with saturation momentum $Q_s$ 
( $\alpha_s= g^2/4\pi \simeq 1/4\pi$ with gauge coupling constant $g\simeq 1$ ),
while the width of the flux tube $R$ is given by $Q_s^{-1}$. 
Hence, we find that the color electric field
decays rapidly such as $t_c\simeq Q_s^{-1} \simeq 0.1\,\rm{fm/c} \sim 0.2\,\rm{fm/c}$
for $Q_s=1\,\rm{GeV}\sim 2\,\rm{GeV}$. The smallness of the width is the main
origin of the rapid decay. 
Although this is the result obtained in the limit of $B\gg E$,
the result indicates that the color electric field decays sufficiently fast
as required phenomenologically\cite{hirano}. In addition,
the result is consistent with our assumption of
the independence of $B$ on time. 
Indeed, $B$ decays very slowly\cite{venugopalan,lappi} 
and does not vary so much at least within the period of $Q_s^{-1}$.

\vspace*{0.3cm}
To summarize, using the chiral anomaly we have 
discussed the pair creation of electrons and positrons in an axial symmetric electric flux tube
under the effect of strong magnetic field.
 We have analytically obtained
the spatial and temporal behaviors of the electric field
and azimuthal magnetic field induced by the current of the fermions.
Our calculations have been performed 
without addressing explicit forms of particle's wavefunctions.
However, our results hold only in the limit of $B\gg E$, in which
all contributions of ``massive" states vanish. 
Applying our results to the glasma produced in high energy heavy-ion collisions, 
we have found that the color electric field decay sufficiently fast to be consistent
with QGP phenomenology.

\hspace*{1cm}

The author
express thanks to Drs. H. Fujii of University of Tokyo and Dr. K. Itakura
of KEK for their useful discussion and comments.



\begin{thebibliography}{99}
\bibitem{tanji}N. Tanji, hep-ph/0810.4429; see the references therein.
\bibitem{schwinger}J. Schwinger, Phys. Rev. {\bf 82} (1951) 664.
\bibitem{iwazaki}A. Casher, H. Neuberger and S. Nussinov, Phys. Rev. {\bf D\,20} (1979) 179.\\
K. Kajantie and T. Matsui, Phys. Lett. {\bf 164\,B} (1985) 373.\\
M. Gyulassy and A. Iwazaki, Phys. Lett. {\bf 165\,B} (1985) 157.
\bibitem{colorglass}
E. Iancu, A. Leonidov and L. McLerran, hep-ph/0202270.\\
E. Iancu and R. Venugopalan, hep-ph/0303204.\\
For a brief review, see K.~Itakura,
 Prog.\ Theor.\ Phys.\ Suppl.\  {\bf 168} (2007) 295.
\bibitem{iwa}A. Iwazaki, Phys. Rev. {\bf C\,77} (2008) 034907; hep-ph/0803.0188, to be published in Prog. Theor. Phys.
\bibitem{itakura}H. Fujii and K. Itakura, Nucl. Phys. {\bf A\,809} (2008) 88.
\bibitem{hii}H. Fujii, K. Itakura and A. Iwazaki, hep-ph/0903.2930.
\bibitem{venugopalan}P. Romatschke and R. Venugopalan, Phys. Rev. Lett. {\bf 96} (2006) 062302;
Phys. Rev. {\bf D\,74} (2006) 045011.
\bibitem{berges}J. Berges, S. Scheffler and D. Sexty, Phys. Rev. {\bf D\,77} (2008) 034504.
\bibitem{iwa2}A. Iwazaki, hep-ph/0904.1449.
\bibitem{hirano}T. Hirano and Y. Nara, Nucl. Phys. {\bf A\,743} (2004) 305; J. Phys. {\bf G\,30} (2004) S1139.
\bibitem{lappi}T. Lappi and L. McLerran, Nucl. Phys. {\bf A\,772} (2006) 200.
\end{thebibliography}
\end{document}